\documentclass[sigconf]{acmart}

\AtBeginDocument{%
  \providecommand\BibTeX{{%
    \normalfont B\kern-0.5em{\scshape i\kern-0.25em b}\kern-0.8em\TeX}}}

\setcopyright{acmcopyright}
\copyrightyear{2022}
\acmYear{2022}
\acmConference[EASE 2022]{The International Conference on Evaluation and Assessment in Software Engineering }{June 13–15, 2022}{Gothenburg, Sweden}
\acmPrice{15.00}
\acmBooktitle {}
\acmISBN {}

\usepackage{algorithmic}
\usepackage{graphicx}
\usepackage{flushend}
\usepackage{dblfloatfix}
\usepackage{textcomp}
\usepackage{xcolor}
\usepackage{url}
\usepackage{booktabs}
\usepackage{tikz}
\usepackage{caption}
\usepackage{subcaption}
\usepackage{multirow}
\usepackage{balance}
\usepackage{colortbl}
\usepackage{listings}
\usepackage{pgfplots}
\usepackage{enumitem}
\usepackage{tabularx}
\usepackage{xurl}

\begin{document}
\renewcommand{\shortauthors}{Ikegami et. al.}

\newcommand\todoV[2]{ {\colorbox{pink}{\textcolor{red}{#1}}} {\todo[color=green!40]{\footnotesize{\thesubsection. #2}}}}
\newcommand\todoU[2]{ {\colorbox{yellow}{\textcolor{red}{#1}}} {\todo[color=red!40]{\thesubsection. #2}}}
\newcommand\todoL[2]{ {\colorbox{yellow}{\textcolor{red}{#1}}} {\todo[color=blue!40]{\thesubsection. #2}}}

\newcommand\greenbox[2][]{\tikz[baseline=(char.base)]\node[minimum width=2em,text height=1.5ex, text depth=0.1ex,fill=green!25,text=black,#1](char){#2};}%
\newcommand\yellowbox[2][]{\tikz[baseline=(char.base)]\node[minimum width=2em,text height=1.5ex, text depth=0.1ex,fill=yellow!25,text=black,#1](char){#2};}%
\newcommand\redbox[2][]{\tikz[baseline=(char.base)]\node[minimum width=2em,text height=1.5ex, text depth=0.1ex,fill=red!25,text=black,#1](char){#2};}%

\newcommand{\RqOne}{\textit{RQ1}}
\newcommand{\RqTwo}{\textit{RQ2}}
\newcommand{\RqOneOne}{\textit{RQ1.1}}
\newcommand{\RqOneTwo}{\textit{RQ1.2}}
\newcommand{\RqOneThree}{\textit{RQ1.3}}
\newcommand{\RqTwoOne}{\textit{RQ2.1}}
\newcommand{\RqTwoTwo}{\textit{RQ2.2}}
\newcommand{\RqThree}{\textit{RQ3}}
\newcommand{\RqFour}{\textit{RQ4}}
\newcommand{\RqFive}{\textit{RQ5}}
\newcommand{\vFR}{secure-fix refactoring}
\newcommand{\VFR}{Secure-fix refactoring}
\newcommand{\vFix}{security-fix}
\newcommand{\vCat}{security}
\newcommand{\vCats}{security categories}
\newcommand{\vOP}{secure-fix refactoring operation}
\newcommand{\vROP}{Secure-fix refactoring operation}
\newcommand{\gRef}{regular refactoring}
\newcommand{\GRef}{Regular refactoring}
\newcommand{\RqOdds}{What is the association between a refactoring type and whether is it used to refactor a security-fix?}
\newcommand{\hypoZero}{\textit{\uline{H$_0$}: \vFR~is same distributed as \gRef}}
\newcommand{\hypoOne}{\textit{\uline{Hypothesis$_{1}$:}
A specific refactoring pattern has is likely to be used to refactor a security-fix.
}}
\newcommand{\hypoTwo}{\textit{\uline{Hypothesis$_{2}$:}
Specific refactoring types are more likely to be used refactor certain vulnerability fixes related to a \CWE.
}}

\newcommand{\hypoThree}{\textit{\uline{Hypothesis$_3$:} 
Developers acknowledge and document the security issue when performing the refactoring.
}}
\newcommand{\hypoFour}{ \textit{\uline{Hypothesis$_4$:} 
Code attributes of the refactoring are likely to be related to the security weakness (CWE) that it fixes.
}}

\title{On the Use of Refactoring in Security Vulnerability Fixes: \\ An Exploratory Study on Maven Libraries}

\author{Ayano Ikegami, Raula Gaikovina Kula}
\affiliation{%
  \institution{Nara Institute of Science and Technology}
  \country{Japan}
}
\email{ikegami.ayano.hs9@is.naist.jp}
\email{raula-k@is.naist.jp}

\author{Bodin Chinthanet, Vittunyuta Maeprasart}
\affiliation{%
  \institution{Nara Institute of Science and Technology}
  \country{Japan}
}
\email{maeprasart.vittunyuta.mn2@is.naist.jp} \email{bodin.ch@is.naist.jp}

\author{Ali Ouni}
\affiliation{%
  \institution{ETS Montreal, University of Quebec, QC}
  \country{Canada}
}
\email{ali.ouni@etsmtl.ca}

\author{Takashi Ishio, Kenichi Matsumoto }
\affiliation{%
  \institution{Nara Institute of Science and Technology}
  \country{Japan}
}
\email{ishio@is.naist.jp}
\email{matumoto@is.naist.jp}

\begin{abstract}
Third-party library dependencies are commonplace in today's software development. 
With the growing threat of security vulnerabilities, applying security fixes in a timely manner is important to protect software systems.
As such, the community developed a list of software and hardware weakness known as Common Weakness Enumeration (CWE) to assess vulnerabilities.
Prior work has revealed that maintenance activities such as refactoring code potentially correlate with security-related aspects in the source code.
In this work, we explore the relationship between refactoring and security by analyzing refactoring actions performed jointly with vulnerability fixes in practice.
We conducted a case study to analyze 143 maven libraries in which 351 known vulnerabilities had been detected and fixed.
Surprisingly, our exploratory results show that developers incorporate refactoring operations in their fixes, with 31.9\% (112 out of 351) of the vulnerabilities paired with refactoring actions. 
We envision this short paper to open up potential new directions to motivate automated tool support, allowing developers to deliver fixes faster, while maintaining their code.
\end{abstract}

\maketitle

\section{Introduction}
\label{sec:intro}
Much like any software, third-party software libraries constantly evolve to meet the needs of clients, patch bugs or address other maintenance concerns \cite{KULA2018186}.
Developers often use refactoring, as an effective practice to maintain their evolving software systems, by restructuring code while keeping its external behaviour \cite{Fowler1999}.
Fowler recommends a set of refactoring actions to improve readability, reusability, and increase the speed to maintain code. 

The threat of security vulnerabilities, especially in library dependencies, is a growing concern.
Developers are wary of the risks of known security vulnerabilities that affect their dependencies \cite{Web:octverse}. 
Security advisories allow maintainers and users to be notified of any potential malicious exploits (identified with a unique CVE identifier \cite{Web:cve}), which is then assessed by a Common Weakness Enumeration (CWE) \cite{Web:cwe}. CWE is a community-developed list of software and hardware weakness types, that serves as a baseline for weakness identification, mitigation, and prevention efforts.
To reduce lags in the delivery of these fixes \cite{Bodin:EMSE2021,Alfadel:SANER2021}, recent advancements in technologies such as bots (i.e., Dependabot \cite{Web:github_dependabot}) are being proposed, so that software teams can quickly fix their software. 
The urgency to respond and fix vulnerabilities has been further amplified with the recent advisories of highly severe security flaws in the third-party libraries such as Log4Shell, affecting thousands of software application that use the Maven Log4J library.\footnote{ An example of the effect can be seen through this gihub repo \url{https://github.com/NCSC-NL/log4shell}}

\begin{figure*}
    \centering
    \includegraphics[width=0.7\textwidth]{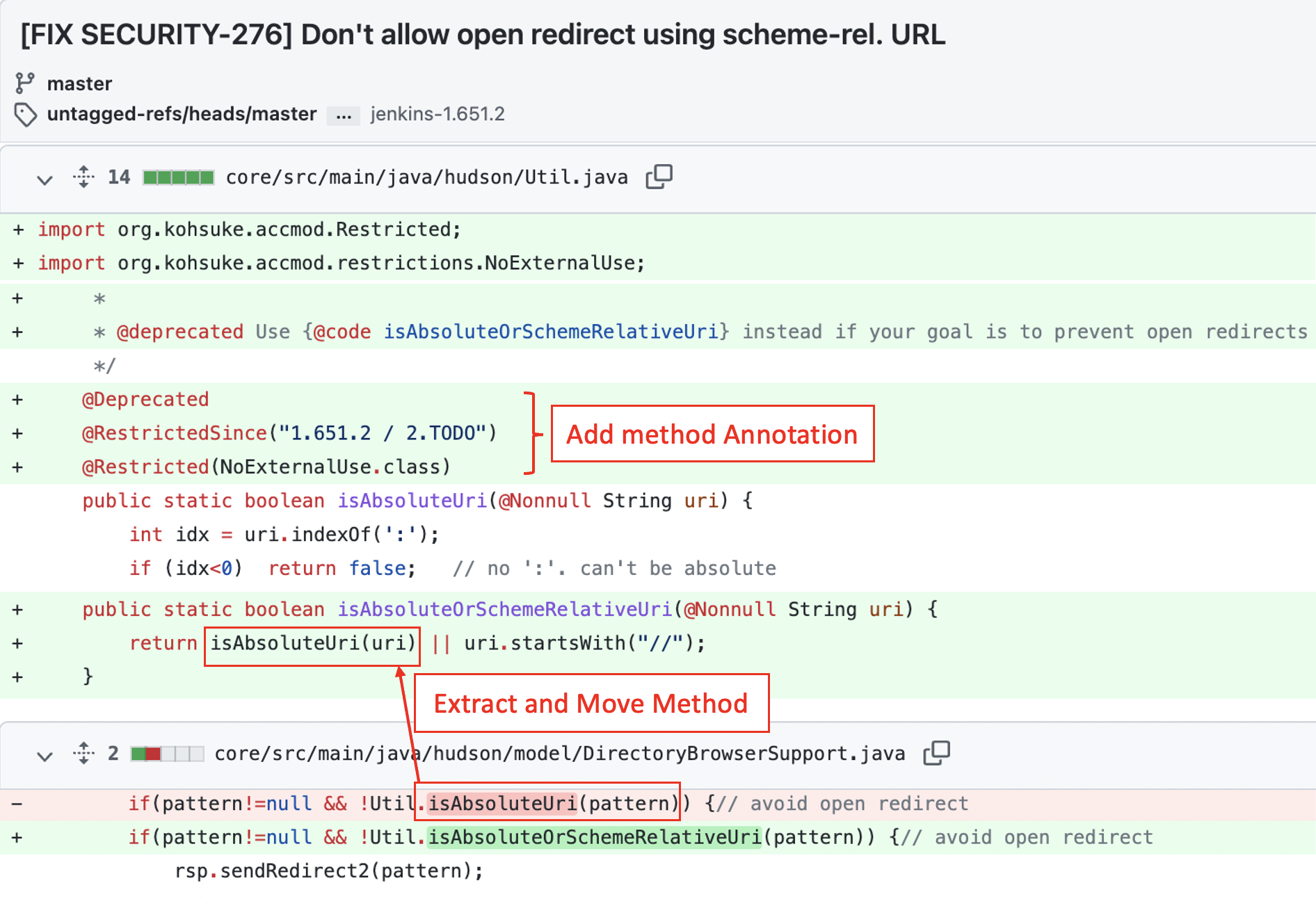}
    \caption{An Example of a Refactor-In-Fix \cite{Web:jenkins2ed}, where the refactoring operation(s) and the vulnerability fix are located in the same commit.}
    \label{fig:case1}
\end{figure*}

Prior work revealed that maintenance activities such as refactoring actions (aka operations) correlate with security-related aspects of code \cite{Abid:TSE2020,MUMTAZ:IST2018}.
For instance, \citet{Abid:TSE2020} found correlations between refactoring operations and security-related aspects such as data-access security vulnerability metrics. Their study reveals that refactoring actions have the potential to impact the security of software systems. On the other hand, \citet{MUMTAZ:IST2018} investigated security metrics (i.e., composition, coupling, extensibility, inheritance and design size) and found that refactoring improves the security of an application without compromising the overall quality in terms of code smells. 
The key motivation that this early research results is to understand the co-relationship between refactoring when fixing a vulnerability.
We hypothesize that fixing a vulnerability may also include cleaning up code.

To fill this gap, in this study, we explore the relationship between refactoring and security vulnerability fixes in practice. 
In particular, we mine and extract refactoring operations that are jointly performed during a vulnerability fix at the same commit (i.e., Refactor-In-Fix). 
Henceforth, we define the following research questions to guide our study:

\begin{itemize}
\item \textbf{RQ1: \textit{How prevalent is refactoring in security vulnerability fixing?}}\\
\textit{Motivation:}
Our motivation is to understand the extent to which refactoring operations are applied to security vulnerability fixes. 
Different to related work, we investigate refactoring-like actions applied when fixing reported security vulnerabilities (CWEs).\\
\textit{Results:}
We find 31.9\% of the security vulnerabilities incorporated refactoring operations in the fix (i.e., 112 vulnerabilities instances).
Furthermore, the \texttt{Extract Method} is most prevalent refactoring type incorporated in a vulnerability fix.
\item\textbf{RQ2: \textit{What refactoring operations do developers apply in security vulnerability fixes?}} \\
\textit{Motivation:} Based on RQ1 results, we
would like to understand which common refactoring patterns that have a higher chance to be applied to refactor a vulnerability fix compared to being applied in other maintenance tasks. \\
\textit{Results:} We identified that the two refactoring types likely incorporated a vulnerability fix are the
 \texttt{Extract Variable} and the \texttt{Extract Method}.
\end{itemize}
\begin{figure*}
    \centering
    \includegraphics[width=0.7\textwidth]{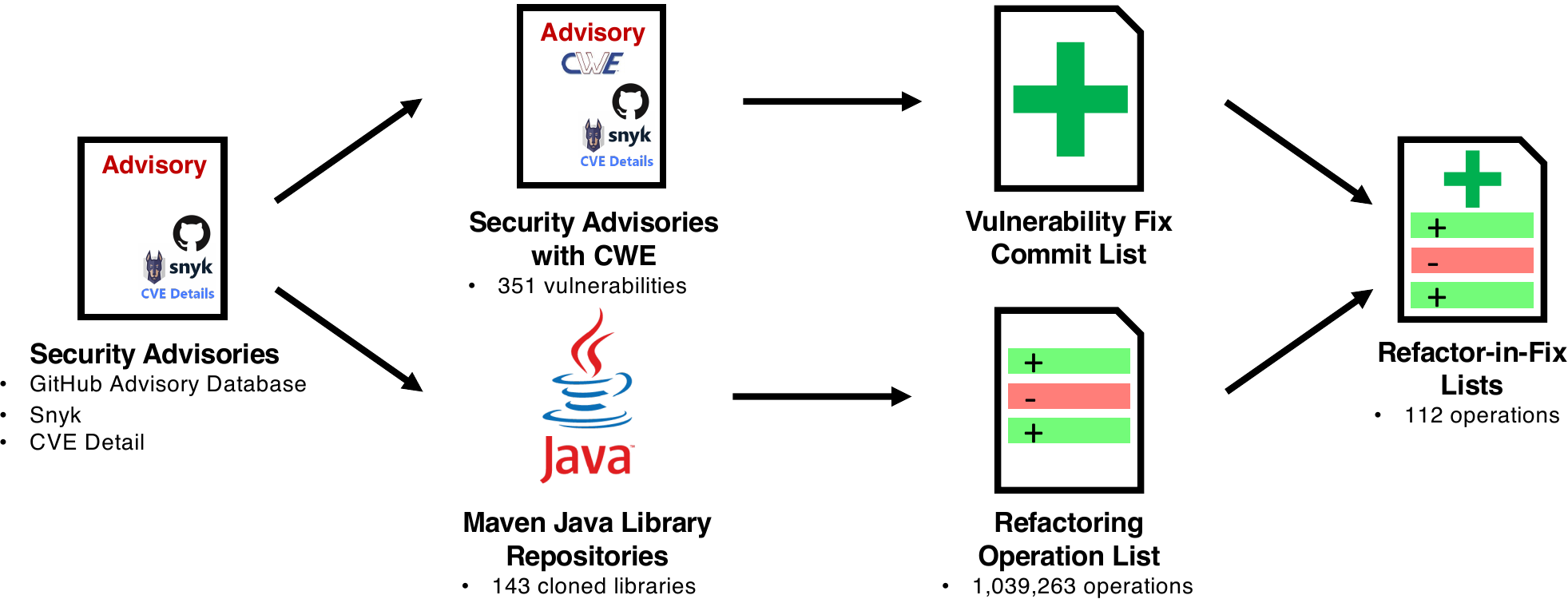}
    \caption{An overview of the data collection process.}
    \label{fig:data_collection}
\end{figure*}
Our results show that there is potential in dedicated refactoring-aware security vulnerability fix techniques, as we believe the adoption of refactoring in security fixes is crucial, especially since it helps developers deliver quicker patches that should complement existing bots like Dependabot \citep{Web:github_dependabot}, thus enabling a faster advisory and mitigation process.

\begin{table}[b]
\centering
\caption{A summary statistic of our data collection.}

\label{tab:dataset_information}

\begin{tabular}{@{}lr@{}}
\toprule
\multicolumn{2}{c}{\textbf{Maven Java Libraries}} \\ \midrule
Repository snapshot & Aug 28, 2020 \\
\# Collected vulnerabilities &  1,809 \\
\# Collected vulnerabilities with CWE & 351 \\
\hspace{2em} - \# Paired with Refactor-In-Fix & 112\\
\hspace{2em} - \# Not paired with Refactor-In-Fix & 239\\
\#  Maven Java libraries identified & 463 \\
\hspace{2em} - \# Cloned Maven Java libraries & 143 \\\hline
\# Refactoring Operations Extracted & 1,039,263 \\
\hspace{2em} - \# Refactor-In-Fix operations & 527 \\
\bottomrule
\end{tabular}
\end{table}

\begin{table*}[]
\centering
\caption{Top 10 most frequent vulnerabilities grouped by CWEs.}
\label{tab:vulnerability}
\scalebox{0.9}{
\begin{tabular}{llrrr}
\hline
\multirow{1}{*}{CWE ID} & \multicolumn{1}{c}{\multirow{1}{*}{CWE Title}  }  
& \multicolumn{1}{c}{\begin{tabular}[c]{@{}l@{}}\#  Vulnerabilities Paired \end{tabular} } &
\multicolumn{1}{c}{\begin{tabular}[c]{@{}l@{}}
     \#  Not Paired
\end{tabular} }
& Total 
           
\\  \midrule

CWE-200 &
  \begin{tabular}{l}Improper Input Validation\end{tabular} &
   $15$ ($34.1\%$) &
  $29$ ($65.9\%$) &
  $44$
   \\
CWE-79 &
  \begin{tabular}{l}Improper Neutralization of Input During Web Page Generation ('Cross-site Scripting')
  \end{tabular}  &
  $4$ ($12.5\%$) &
  $28$ ($87.5\%$)&
  $32$ 

 \\
      CWE-502 &
 \begin{tabular}{l}
      Deserialization of Untrusted Data
 \end{tabular} &
   $5$ ($19.2\%$) &
   $21$ ($80.8\%$) &
   $26$ 

   \\
   \rowcolor{green!25}CWE-611 & \begin{tabular}{l}Improper Restriction of XML External Entity Reference\end{tabular} & $16$ ($66.7\%$) & $8$ ($33.3\%$) & $24$

\\
 CWE-284 &
 \begin{tabular}{l}
       Improper Access Control
 \end{tabular}
    &
   $1$ ($5.3\%$) &
   $18$ ($94.7\%$) &
   $19$

    \\
     
 \rowcolor{green!25} CWE-94 &
  \begin{tabular}{l}Improper Control of Generation of Code ('Code Injection')\end{tabular} &
   $10$ ($55.6\%$) &
   $8$ ($44.4\%$) &
   $18$ 

  \\
  CWE-22 &
  \begin{tabular}{l}
  Improper Limitation of a Pathname to 
  a Restricted Directory ('Path Traversal')     
  \end{tabular}
   &
   $1$ ($5.9\%$)  &
   $16$ ($94.1\%$) &
   $17$

   \\
  \rowcolor{green!25} CWE-352 &
 \begin{tabular}{l}
      Cross-Site Request Forgery (CSRF)
 \end{tabular} &
   $9$ ($60.0\%$) &
   $6$ ($40.0\%$) &
   $15$ 

   \\

 CWE-264 &
  \begin{tabular}{l}
Permissions, Privileges, and Access Controls
  \end{tabular}
    &
   $4$ ($30.8\%$) &
   $9$ ($69.2\%$) &
   $13$ 

    \\
         CWE-400 &
 \begin{tabular}{l}
      Uncontrolled Resource Consumption
 \end{tabular} &
   $4$ ($36.4\%$) &
   $7$ ($63.6\%$) &
   $11$
   \\ 
   \multicolumn{1}{c}{...} & \multicolumn{1}{c}{...} & \multicolumn{1}{c}{...} & \multicolumn{1}{c}{...} & \multicolumn{1}{c}{...}\\
   
   \midrule
    &
  \multicolumn{1}{r}{Total} &
  $112$ ($31.9\%$) & $239$ ($68.1\%$) & $351$ 
    \\ \hline
\end{tabular}
}
\end{table*}
\section{Data collection}
In this section, we present our approach to pair a refactoring operation to a vulnerability fix, and then our dataset used in the study.

\paragraph{\textbf{Pairing Refactoring Operations With a Vulnerability Fix (Refactor-In-Fix)}}

A key challenge to respond our research questions, is to accurately identify which refactoring operations were applied to a vulnerability fix.
To ensure relateness, we pair any refactoring operations that appear in the same commit as the vulnerability fix.

Figure \ref{fig:case1} shows an example of a refactoring operation  taken from the Jenkins project \cite{Web:jenkins}.
In this case, the \texttt{Extract and Move Method} is being applied in the same commit as the vulnerability fix \cite{Web:jenkins2ed}. 
The security vulnerability being fixed is related to \textit{`Open redirect to scheme-relative URLs'} (\texttt{CVE-2016-3726}) \cite{Web:NVD2016}. 
Furthermore, the CWE is related to \textit{URL Redirection to Untrusted Site} (\texttt{CWE-601}).
This is a severe weakness, with a high score (7.4 HIGH) ranking according to the National Vulnerability Database \cite{Web:NVDHome}.

Interestingly, we can also see that the vulnerability fix includes the \texttt{Add Method Annotation} and the \texttt{Extract and Move Method} refactoring operations.
More specifically, developers added the new annotations \texttt{@Deprecated,} \sloppy{\texttt{@Restricted, @RestrictedSince}}~to show deprecations of the method. 
We also see developers performed an extract and move of the \texttt{.isAbsoluteUri} and replace it with a \texttt{isAbsoluteOrSchemeRelativeUri} method to secure the code.

\begin{table}[t]
\centering
\caption{Top 10 most frequent refactoring types paired to a vulnerability fix}
\label{tab:refactoring_top}
\begin{tabular}{@{}lcr@{}}
\hline
\multirow{1}{*}{Refactoring type} &
  \multicolumn{1}{l}{\begin{tabular}[c]{@{}l@{}} \# Refactor-In-Fix\end{tabular}}
  & \# Other

   \\ \midrule
Extract Method                                 & $86$  
& $37,688$
 
\\
Add Method Annotation &                  $69$
& $177,153$

\\
Rename Method         &                  $58$ &
$52,622$

\\
Extract Variable                              & $51$
        & $19,249$

\\
Rename Parameter                               & $36$  
        & $43,927$

\\
Change Variable Type                          & $33$ 
        & $87,651$

\\
Rename Variable                                & $27$  
& $43,982$

\\
Change Attribute Type                         & $27$  
& $42,077$

\\
Rename Attribute                              & $20$
& $19,717$

\\
Change Return Type    & $18$                        & $58,939$
\\
\multicolumn{1}{c}{...} & \multicolumn{1}{c}{...} & \multicolumn{1}{c}{...}
\\ \midrule
    \multicolumn{1}{r}{Total} & $527$ & $1,039,263$ 

      \\ \hline
\end{tabular}

\end{table}

\paragraph{\textbf{Mining Maven Libraries}}
\label{sec:preliminary}
Our approach is to mine software repositories to detect refactorings applied on vulnerability fixes.
Since we wanted to explore this phenomenon in the wild, we selected Java libraries from Maven \citep{MavenCentral:online}, one of the largest third-party library ecosystems.
The Maven ecosystem of libraries is well-known and actively reports and fixes vulnerabilities, as shown by the number of vulnerabilities that appear in the GitHub Advisory Database \citep{Web:github:advisories}. 
Figure \ref{fig:data_collection} shows the overview of our data collection and how we process them.
We first needed to identify libraries that had reported and fixed a security vulnerability. 
Following prior works \cite{Bodin:EMSE2021,Li:CCS2017,Perl:CCS2015,Du:ICSE2019,Fan:MSR2020}, we collected data from three reputable data sources (\textit{i}) GitHub Advisory Database \cite{Web:github:advisories}  (\textit{ii}) Snyk \cite{Web:snyk} and (\textit{iii}) CVE Details \cite{Web:cvedetail}.
Similar to \cite{Bodin:EMSE2021,Alfadel:SANER2021}, we then crawl the advisories to extract any linkages to a GitHub commit, and has a CWE associated with the vulnerability.
In the end, from an initial list of vulnerabilities that contains 1,809 advisories, we identified 463 vulnerable Java libraries. 
Since we were only able to clone 143 repositories, as they either had accessibility issues (permissions needed) or that the repository was not longer available, we ended up with 351 vulnerabilities.

To identify and extract the applied refactoring operations, we use the RefactoringMiner tool \cite{RMiner,Tsantalis:ICSE2018}, as it is the
state-of-the-art for detecting refactoring operations in Java programs, and is widely used in the community \cite{related_secure_ref_Di,related_ref_alomar,hamdi2021longitudinal}. 
From the 143 cloned repositories, we extract 1,039,263 refactoring operations. 
In the end, we paired refactoring operations with vulnerability fixes in the same commit (as Refactor-In-Fix).
A summary statistic of the data collection is shown in Table \ref{tab:dataset_information}.

\section{Exploratory Study Results}
We now present the results of our study.

\subsection{Prevalence Analysis (RQ1)}
To answer RQ1, we grouped our collected dataset by CWEs as shown in Table \ref{tab:vulnerability}, and refactoring types in Table \ref{tab:refactoring_top}. 
We make two main observations.

First,  we find that 31.9\% of the security vulnerabilities incorporated refactoring operations in the fix.
Additionally, as shown in Table \ref{tab:vulnerability}, we find three security weaknesses
(i.e., \texttt{CWE-611} \cite{Web:CWE-611}, \texttt{CWE-94} \cite{Web:CWE-94}, and  
\texttt{CWE-352} \cite{Web:CWE-352})
that have more vulnerabilities paired that those vulnerabilities that are not paired.
In detail, \texttt{CWE-611} \textit{allows an attacker to access arbitrary files on the system}, while \texttt{CWE-94} includes \textit{Code injection attacks can lead to loss of data integrity in nearly all cases as the control-plane data injected is always incidental to data recall or writing, as well as result in the execution of arbitrary code.}
Finally, \texttt{CWE-352} is a weakness \textit{where an attacker could effectively perform any operations as the victim. If the victim is an administrator or privileged user, the consequences may include obtaining complete control over the web application - deleting or stealing data, uninstalling the product}.
Second, as shown in Table \ref{tab:refactoring_top}, we find that the \texttt{Extract Method} is the most prevalent refactoring type applied in a vulnerability fix.
Conversely, the \texttt{Add Method Annotation} is the most common refactoring type that is applied to other commits (not on the vulnerability fix) in the source code.

\begin{quote}
\textbf{Summary}: To answer RQ1, we find that 31.9\% of the security vulnerabilities had incorporated refactoring operations in the fix (i.e., 112 vulnerabilities instances).
We also find that the \textit{Extract Method} is most prevalent refactoring type incorporated in a vulnerability fix.
\end{quote}

\subsection{Refactoring operations associated with vulnerability fixes (RQ2)}
To answer RQ2, we use a statistical measure to understand the likelihood that a refactoring operation is performed in a vulnerability fix. 
Similar to \citet{Martina2019}, we use the odds ratio \cite{odds} as a “measure of association”, to quantify this relationship.
Using the results from RQ1, we calculate the top 10 most frequent refactoring types (527 refactoring operations paired with vulnerabilities and 1,039,263 refactorings from other commits in the code, as shown in Table \ref{tab:refactoring_top}).
For our statistical analysis, we test the null hypothesis that:
\textit{{H$_{1}$}: The odds of using a certain refactoring operation to refactor a vulnerability fix is strong, }where the \textit{odds ratio} refers to the strength of the relationship, and whether it is likely to be applied to a vulnerability fix (ratio is $>$ 1) or applied as a regular refactoring (ratio is $<$ 1).
To statistically validate our hypothesis, we calculated the odds ratio using a 95\% confidence interval and use the chi-squared test (\textit{p-value} $<$ 0.05).
We use the online odds ratio calculator.\footnote{\url{https://www.medcalc.org/calc/odds_ratio.php}} 

\begin{table}[t]
\caption{Proportion of the refactoring operations being applied to a vulnerability fix vs. others: Chi-square test p-values and odds ratio. ($\ast$: p-value $<$ 0.05) }
\label{tab:odds_in_fix}
\begin{tabular}{@{}lr@{}}
\toprule
Refactoring Type     &
\multicolumn{1}{l}{Odds Ratio} 
\\ \midrule 
Extract Variable                 & $5.678^{\ast}$   
\\
Extract Method                    & $5.157^{\ast}$  
\\
Rename Method                    & $2.319^{\ast}$   
\\
Rename Attribute & $2.040^{\ast}$ 
\\
Rename Parameter     & $1.662^{\ast}$   
\\
Change Attribute Type &
$1.280\phantom{a}$ 
\\
Rename Variable & $1.222\phantom{a}$ 
\\
  \rowcolor{red!15}Add Method Annotation                      & $0.733^{\ast}$ 
\\
\rowcolor{red!15}Change Variable Type &
$0.725\phantom{0}$ 
\\
\rowcolor{red!15}Change Return Type & $0.588^{\ast}$ 
\\
\bottomrule
\end{tabular}
\end{table}

Table \ref{tab:odds_in_fix} reports our findings, from which we make two main observations. 
First, the two most likely refactoring types to be applied to a vulnerability fix are
the \texttt{Extract Variable} (odds ratio of 5.678) and the \texttt{Extract Method} (odds ratio of 5.157).
This result is consistent with our finding in RQ1 (cf. Table \ref{tab:refactoring_top}).
On the other hand, we find three refactoring types that are more likely to be applied to other commits. 
The three refactoring types are \texttt{Add Method Annotation}  with an odds ratio of 0.733, \texttt{Change Variable Type} with an odds ratio of 0.725, \texttt{Change Return Type} with an odds ratio of 0.588.
We find that except for three cases (i.e., \texttt{Change Attribute Type} and \texttt{Rename Variable}, \texttt{Change Variable Type}), all odds ratios are statistically significant.

\begin{quote}
\textbf{Summary}:
Results show that the two most likely refactoring types to be applied to a vulnerability fix is
 \texttt{Extract Variable} (odds ratio of 5.678) and \texttt{Extract Method} (odds ratio of 5.157).
 On the other hand, we identified three refactoring types (i.e., \texttt{Add Method Annotation}, \texttt{Change Variable Type}, \texttt{Change Return Type}) that are more likely to be applied as \textit{regular} refactoring operations. 
\end{quote}

\section{Threats to Validity}
As a short paper, the key threat is the completeness and vigor of the sample size, and a manual validation of refactoring miner shortcomings (i.e., accuracy). 
We do acknowledge that our current pairing is a very conservative approach, as developers may applied refactoring operations in subsequent commits after the vulnerability fix.
However, we argue that our result is substantial and calls attention for researchers to further explore this research avenue.

Another key limitation, we we plan to address in the full study, is to handle vulnerabilities fixed across multiple commits. 
To mitigate this, we will include these vulnerabilities in the future study.

\section{Implications And Future Plans}
Our exploratory study delivers various important takeaways and implications to existing practices, and future plans as our roadmap.

\textbf{Takeaways.}
With recent works exploring the inducing effects of refactoring actions on software maintenance tasks such as bug fixes, Pull Request acceptance and security-aspects \cite{Penta2020,Bavota2012,Abid:TSE2020, Flavia2021, Wanwangying2016, Maruyama:ICSOFT2007}, we take a reverse approach to explore further how refactoring is used to help in vulnerability fixes. 
Our key takeaway message as highlighted in RQ1,  is that developers do incorporate refactoring (i.e., 31.9\%) operations when applying a vulnerability fix in practice.
Furthermore, as shown in RQ2, we distinguish with statistical significance, the odds that a certain refactoring type is paired with a vulnerability fix. 
This potentially spark research into more dedicated refactoring-aware security vulnerability fix techniques.

\textbf{Disruption to Current Practices.} We believe the adoption of refactoring in security fixes is crucial, especially since it helps developers deliver quicker patches, while maintaining the quality of their code through refactoring. 
Complementing existing bots like Dependabot \citep{Web:github_dependabot}, automated tooling may help developers understand how vulnerabilities could be fixed in the future, thus improving the vulnerability advisory process.
As shown by recent studies, there are lags and issues that exist with fixing vulnerabilities \citep{Bodin:EMSE2021, Zerouali:ICSR2018, Alfadel:SANER2021, Decan:ICSME:2018, Kikas:2017}. 

\textbf{Future Plans.}
Motivated by our quantitative results, we plan to execute a more comprehensive study on Java applications, as well as Maven libraries and other library ecosystems.
Furthermore, we would like to extend our pairing approach and use other heuristics such as the commit messages and the location at the lines-of-code granularity to achieve more accurate results.
We plan our extended study to also include a qualitative component, where we can manually validate how the refactoring types are implemented, and how automation could be achieved. 
Finally, we plan to conduct a pre and post-assessment experiment, using existing code quality metrics and code smells to validate whether or not vulnerability fixes paired with refactoring operations do impact the code quality. 

\begin{acks}
This work has been supported by Japan Society for the Promotion of Science KAKENHI Grants Grant Number JP20H05706 and JP20K19774.
Ali is supported by Natural Sciences and Engineering Research Council of Canada (NSERC) discovery grant RGPIN-2018-05960.
\end{acks}

\balance
\bibliographystyle{ACM-Reference-Format}
\bibliography{sample-base}


\end{document}